\documentclass[twocolumn,showpacs,superscriptaddress,preprintnumbers,amsmath,amssymb,prb]{revtex4}

\usepackage{graphicx}
\usepackage{dcolumn}
\usepackage{bm}
\usepackage{ulem}
\usepackage{color}

\begin{document}

\title{
Optical metamaterials with different metals}

\author{Nian-Hai Shen}
\email[]{nhshen@ameslab.gov} \affiliation{Ames Laboratory and
Department of Physics and Astronomy, Iowa State University, Ames,
Iowa 50011, U.S.A.}

\author{Thomas Koschny}
\affiliation{Ames Laboratory and Department of Physics and
Astronomy, Iowa State University, Ames, Iowa 50011, U.S.A.}
\affiliation{Institute of Electronic Structure and Laser, FORTH,
71110 Heraklion, Crete, Greece}

\author{Maria Kafesaki}
\affiliation{Institute of Electronic Structure and Laser, FORTH,
71110 Heraklion, Crete, Greece} \affiliation{Department of Materials
Science and Technology, University of Crete, 71003 Heraklion, Crete,
Greece}

\author{Costas M. Soukoulis}
\email[]{soukoulis@ameslab.gov} \affiliation{Ames Laboratory and
Department of Physics and Astronomy, Iowa State University, Ames,
Iowa 50011, U.S.A.} \affiliation{Institute of Electronic Structure
and Laser, FORTH, 71110 Heraklion, Crete, Greece}

\begin{abstract}
We investigate the influence of different metals on the
electromagnetic response of fishnet metamaterials in the optical
regime. We found, instead of using a Drude model, metals with a
dielectric function from experimentally-measured data should be
applied to correctly predict the behavior of optical metamaterials.
Through comparison of the performance for fishnet metamaterials made
with different metals, i.e., gold, copper, and silver, we found
silver is the best choice for the metallic parts compared to other
metals, because silver allows for the strongest
negative-permeability resonance and, hence, for optical fishnet
metamaterials with a high figure-of-merit. Our study offers a
valuable reference in the designs for optical metamaterials with
optimized properties.
\end{abstract}


\maketitle

Research on metamaterials (MMs), i.e., artificial structures
composed of tailored subwavelength building blocks, has been a
burgeoning field in the past decade, due to great interest in both
theoretical studies and practical applications. \cite{Shalaev2007,
Soukoulis2007, Zheludev2010, Boltasseva, Soukoulis2011} MMs are
found able to greatly improve our capabilities to manipulate
electromagnetic radiation almost throughout the entire spectrum,
providing many intriguing properties and phenomena, such as negative
refractive index $n$,\cite{Shelby2001, Pendry2000}
superlensing,\cite{Pendry2000, Jacob2006, Liu2007, Smolyaninov2007}
and invisibility.\cite{Pendry2006, Schurig2006, Cai2007,
Valentine2009, Gabrielli2009} Following the efforts of pushing MMs
to work in the optical regime \cite{Valentine2008, Garcia2011,
Chanda2011} from the microwave regime,\cite{Shelby2001, Koray2005}
intrinsic loss of constituents, especially metallic inclusions, has
become a more severe problem. In the course of realizing negative
$n$ in the visible regime, fishnet, i.e., a perforated
metal-dielectric-metal sandwich, has substituted
split-ring-resonator (SRR), which fails to provide negative $\mu$
above infrared frequencies,\cite{Zhou2005} to become the most
promising structure. Concerning strategies for reducing loss in a MM
system, an appropriate selection of low loss materials as
constituents of MMs is quite straightforward. The introduction of
gain materials is believed to be the most efficient but also very
challenging.\cite{Xiao2010, Soukoulis2010} For pure passive
structures, lower losses have been reported by building a
strongly-coupled system \cite{Zhou2009, Valentine2008} and by
geometric optimizations.\cite{Durdu2009} In the meantime, because of
costly fabrication of MMs in the optical regime, guidance from
simulations is helpful and desirable. However, it is noted all the
related materials should be correctly modeled to accurately predict
the behaviors of MMs. Here, we focus on the influence of metals in
the electromagnetic response of MMs, show how we should model metals
correctly, and which metal is the best choice for achieving low-loss
negative-index MMs in the optical regime.

The discussions in this paper are divided into two parts: First,
taking gold as an example, we compare the experimentally-measured
Johnson-Christy (JC) data \cite{JC1972} with some Drude models in
the optical regime and investigate their influence on the response
of a fishnet MM. Second, we study the performances of a MM in the
visible regime based on different metals to show the best option of
metal as a constituent of MMs to obtain a satisfying negative-$n$
property.

For noble metals, there are various sources providing different sets
of frequency-dependent material data to describe them.\cite{JC1972,
Ordal1987} Because of the additional scattering for electrons
resulting from the metal surfaces, material property of a thin-film
metal is different from that of a bulk metal. JC data, specifically
measured for thin film metals and proven independent of thickness
between 25-50 nm, is very appropriate for application in the
modeling of MMs in the optical regime, in which the thickness of
metallic inclusions is generally tens of nanometers. On the other
hand, the Drude model, i.e., $\varepsilon(f) = \varepsilon_{\infty}
- f_{p}^2/[f(f+if_c)]$, where $f$ is the frequency,
$\varepsilon_{\infty}$ is the offset value of permittivity, while
$f_p$ and $f_c$ are plasma and collision frequencies, respectively,
is also widely applied to describe metals. However, we must be
careful that when our interested frequency band goes into the
optical regime, the Drude model may no longer be valid, as shown
below.

A widely-used Drude model (model 1) \cite{Linden2004} in the field
of MMs for gold has $\varepsilon_{\infty}=1$, $f_p=2175$ THz, and
$f_c=10.725$ THz, with $f_c$ three times larger than the normal
value for bulk metal shown in Ref. [24]. We show Drude model 1 with
the frequency-dependent dielectric constant from 150 to 750 THz
compared to JC data in Fig. 1. We clearly see the discrepancy
between the two for both real and imaginary parts of permittivity
throughout the range. The experimental JC data include higher
Im[$\varepsilon$], thereby a higher intrinsic loss in gold.
Intuitively, the application of Drude model 1 for gold in the design
of MMs, such as fishnet structures (see Fig. 2), may lead to an
inaccurate and overly optimistic prediction of performance. On the
other hand, we have searched for an improved Drude model for gold,
which fits JC data better and found with $\varepsilon_{\infty}=9.6$,
$f_p=2184$ THz, and $f_c=17$ THz, i.e., Drude model 2, agrees with
JC data perfectly up to 450 THz (shown in Fig. 1). However,
thereafter, the goodness of fit between Drude model 2 and JC data
starts breaking for Im[$\varepsilon$]. Therefore, for frequencies
above 450 THz, even Drude model 2 will not be valid to describe
gold's properties.

\begin{figure}
 \centering
 \includegraphics[width=8 cm]{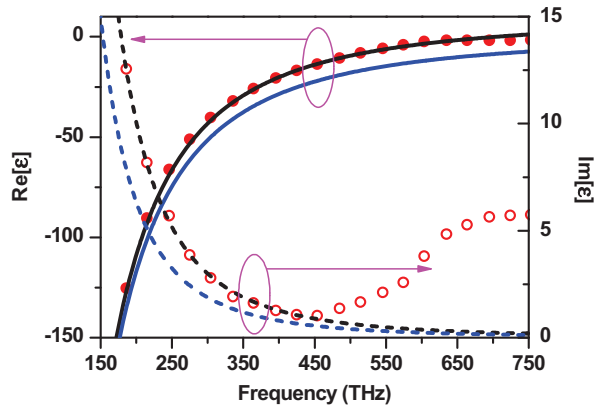}
 \caption{(Color online) Comparison of permittivity (both real
and imaginary parts) for gold with Johnson-Christy data (symbols),
Drude model 1 (blue lines), and Drude model 2 (black lines).}
\end{figure}

Provided the fishnet structure (shown in Fig. 2) is the most popular
MM in the optical regime, due to the ease in fabrication and
satisfying performance in achieving double negative ($\varepsilon<0$
and $\mu<0$) property, we would like to take the fishnet as an
example to show the influence on the electromagnetic response of
MMs, when we apply different sets of material data for metal (gold)
inclusions.

The first specific fishnet structure, geometric parameters presented
in the caption of Fig. 3, has the expected interesting
negative-index property within telecommunications wavelengths
(around 1.5 $\mu$m). For the metal inclusions (gold), we apply,
respectively, JC data, Drude models 1 and 2. On the other hand, the
dielectric spacer is lossless magnesium oxide (MgO) with $n= 1.7$.
We correspondingly show calculated
reflection/transmission/absorption (RTA) information \cite{CST} and
retrieved effective electromagnetic parameters,\cite{Smith2002,
Chen2004, Smith2005, Koschny2005} i.e., $n$, permittivity
($\varepsilon$), and permeability ($\mu$) in columns (a), (b), and
(c), respectively. It should be pointed out that the retrieval
process is not trivial in general, especially when metamaterials are
anisotropic or bi-anisotropic \cite{Ku2009} and EM wave is oblique
incident.\cite{Menzel2010} From Fig. 3, we find that even though all
three cases show us double-negative properties, the results of Case
(b) are moderately superior over the other two cases, i.e., more
promising negative value of $\mu$ and figure-of-merit
(FOM=$\mid$Re[$n$]/Im[$n$]$\mid$) are reached in Fig. 3(b), which
result from a stronger magnetic resonance under the lower-loss
consideration in Drude model 1. In comparison, Cases (a) and (c)
present almost the same results, which are expected because of the
perfect fit of Drude model 2 to JC data for gold within our present
studied regime (see Fig. 1).

In Fig. 4, we show simulated RTA results and retrieved spectra of
$n$, $\varepsilon$, and $\mu$ for the fishnet structure 2,
specifically designed to work in the visible regime. Again, columns
(a), (b), and (c) correspond to the cases with gold modeled by JC
data, Drude models 1 and 2, respectively. From Fig. 4(a), a magnetic
resonance occurs around 450 THz with positive values of $\mu$
throughout the region. Nevertheless, negative $n$ does exist with
quite low values of FOM ($\sim$0.5 at maximum). Because of the
discrepancy for Im[$\varepsilon$] of gold between Drude model 2 and
JC data above 450 THz, the electromagnetic response of the fishnet
shown in Fig. 4(c) is significantly different from that in Fig. 4(a)
for high frequencies. In comparison, Fig. 4(b) still shows
simultaneously negative $\varepsilon$ and $\mu$, which is
nonrealistic. It is determined magnetic resonance occurs at a higher
frequency compared to that in the other two cases.

\begin{figure}
 \centering
 \includegraphics[width=6.5 cm]{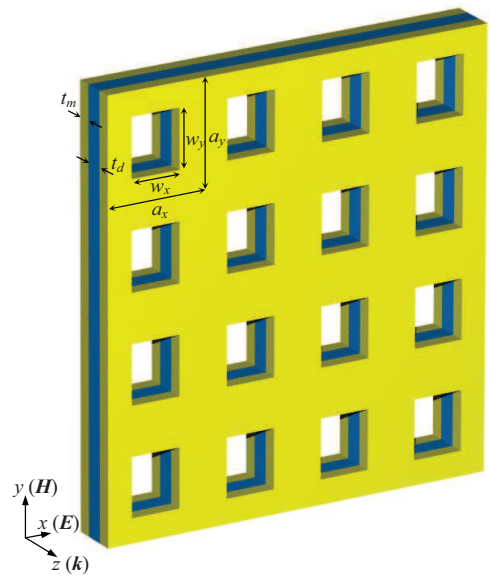}
 \caption{(Color online) Sketch of the fishnet structure: $a_x$ and $a_y$ are
the unit cell sizes along $x$ and $y$ directions, respectively.
$w_x$ and $w_y$ show the hole size. $t_m$ and $t_d$ are the
thicknesses of metallic cladding and dielectric spacer,
respectively.}
\end{figure}

\begin{figure*}
 \centering
 \includegraphics[width=12 cm]{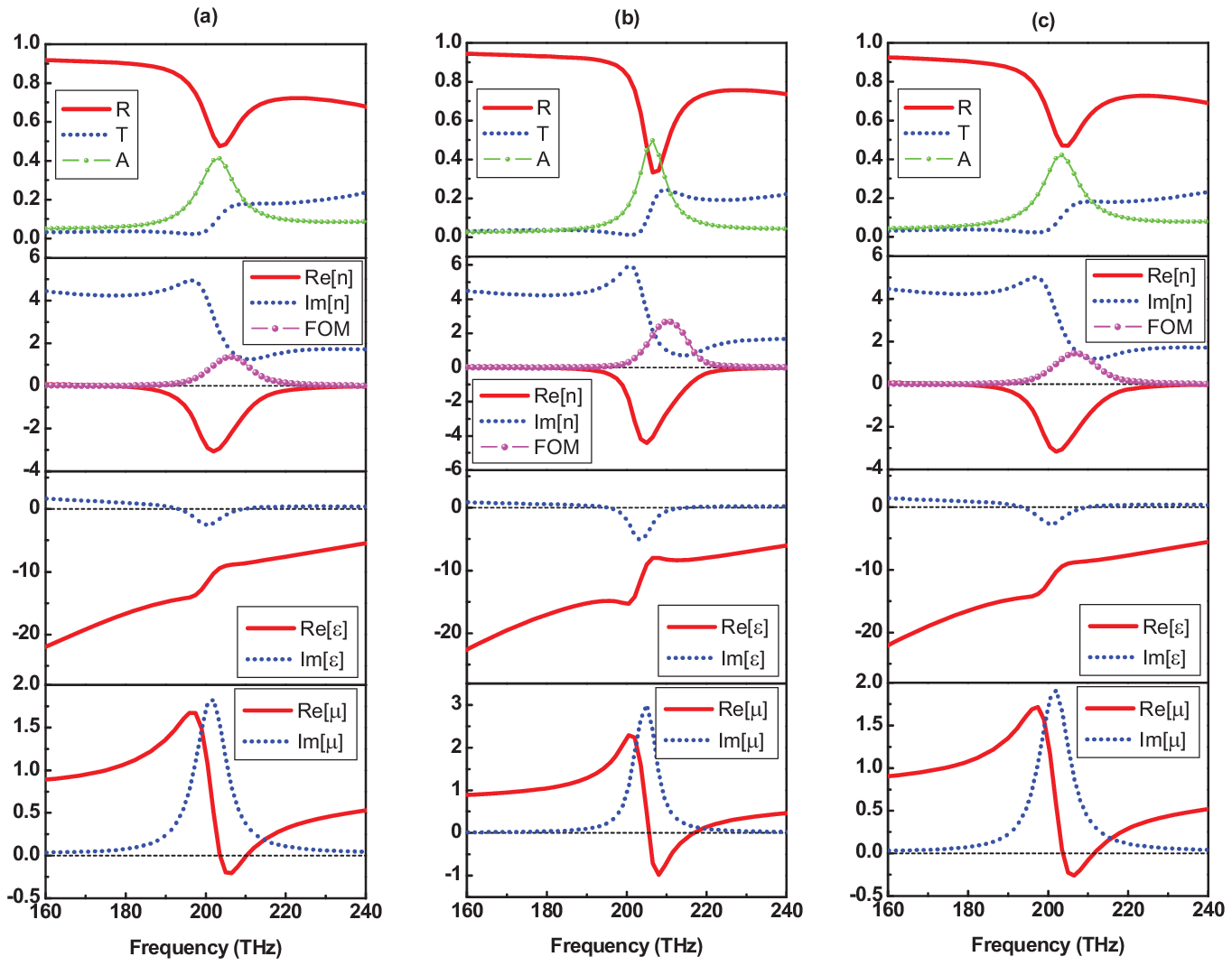}
 \caption{(Color online) Simulated reflection, transmission, absorption, and
retrieved electromagnetic parameters for the fishnet shown in the
inset of Fig. 2 with $a_x=500$ nm, $a_y=600$ nm, $w_x=200$ nm,
$w_y=350$ nm, $t_m=30$ nm, and $t_d=40$ nm for cases in which metal
(gold) is modeled by discrete JC data (a), Drude model 1 (b), and
Drude model 2 (c).}
\end{figure*}

\begin{figure*}
 \centering
 \includegraphics[width=12 cm]{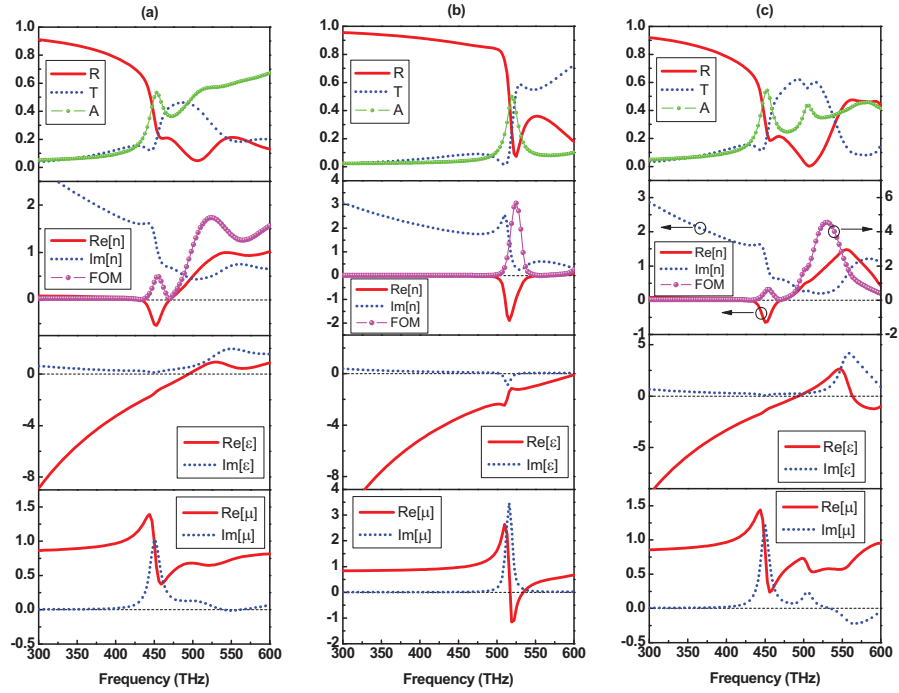}
 \caption{(Color online) Simulated reflection, transmission, absorption, and
retrieved electromagnetic parameters for the fishnet shown in Fig. 2
with $a_x=200$ nm, $a_y=220$ nm, $w_x=100$ nm, $w_y=120$ nm,
$t_m=30$ nm, and $t_d=40$ nm for cases in which metal (gold) is
modeled by discrete JC data (a), Drude model 1 (b), and Drude model
2 (c).}
\end{figure*}

Therefore, we obtain the first important conclusion: to correctly
predict the electromagnetic behavior of MMs in the optical regime
(both telecommunication and visible wavelengths), the metallic
inclusions should be modeled with accurate data, like experimental
JC data. Drude model 1, though widely used, is inaccurate or even
invalid to describe gold in the optical regime and may deceive us to
show the promising negative $\mu$, which is nonrealistic for related
MMs. An improved Drude model 2 fits JC data perfectly up to 450 THz
and provides an alternate for modeling gold, when discrete JC data
cannot be applied, such as in the time domain
calculations.\cite{Drude_Lorentz}

\begin{figure}
 \centering
 \includegraphics[width=6.5 cm]{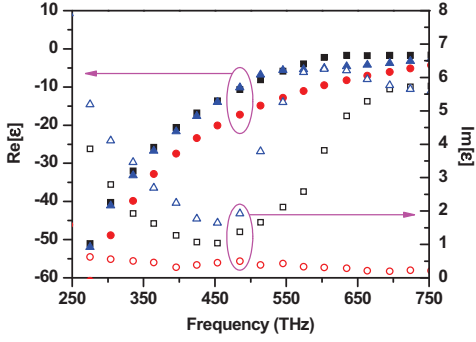}
 \caption{(Color online) Comparison of real (solid symbols) and imaginary (hollow symbols)
 parts of permittivities for gold (square), silver (circle), and copper (triangle).}
\end{figure}

In the following section, we would like to study the influence of
different metals to a metallodielectric MM in the visible regime to
show which metal will be the best choice to benefit MMs in the
achievement of a negative-index property with a high FOM. Based on
our above discussions, to study the properties of MMs in the optical
regime accurately, experimentally-measured material data are always
preferred to apply to the metal inclusions. Figure 5 shows JC data,
i.e., both real and imaginary parts of permittivity from 250 to 750
THz, for thin film of gold, copper, and silver. As seen in Fig. 5,
throughout this frequency range, silver has the lowest loss, so a
silver-made fishnet might be able to render the most promising
property. However, such a speculation needs to be confirmed through
the comparison of the real performances for a MM made with different
metals. Using fishnet structure 2, whose geometric parameters are
shown in the caption of Fig. 4, as an example, we perform
simulations obtaining RTA information and then retrieve the
effective $n$, $\varepsilon$, and $\mu$ of the structure.

\begin{figure}
 \centering
 \includegraphics[width=8 cm]{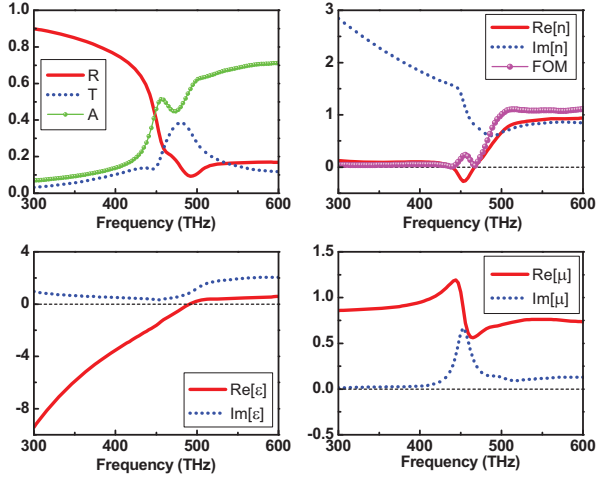}
 \caption{(Color online) Simulated reflection, transmission, absorption, and
retrieved electromagnetic parameters for structure 2, in which,
copper metal is applied with the corresponding JC data.}
\end{figure}

The case for gold metal in the fishnet has been presented in Fig.
4(a): a magnetic resonance occurs at around 450 THz without any
negative $\mu$, but negative $n$ is achieved with a quite low FOM
($\sim$0.5 at maximum). In comparison, the cases for copper- and
silver-made fishnet are shown, respectively, in Figs. (6) and (7).
With copper, we find the performance of the fishnet is even worse
than the case with gold, i.e, FOM ($\sim$0.25 at maximum) is very
low within the narrow band corresponding to negative $n$. The
results for the case with silver are really exciting: Simultaneously
negative $\varepsilon$ and $\mu$ are observed in a narrow band
around 500 THz, leading to a negative $n$ with quite satisfying FOM
(more than 2 at maximum). Note here we also tried the case with
aluminum, the data of which are from Ref.[26] and not JC data.
Results (not shown here) are not as good as those for silver.
Therefore, due to silver's quite low intrinsic loss, silver is
supposed to be the best option for metal inclusions of
metallodielectric MMs to achieve negative index in the visible
regime.

\begin{figure}[htb]
 \centering
 \includegraphics[width=8 cm]{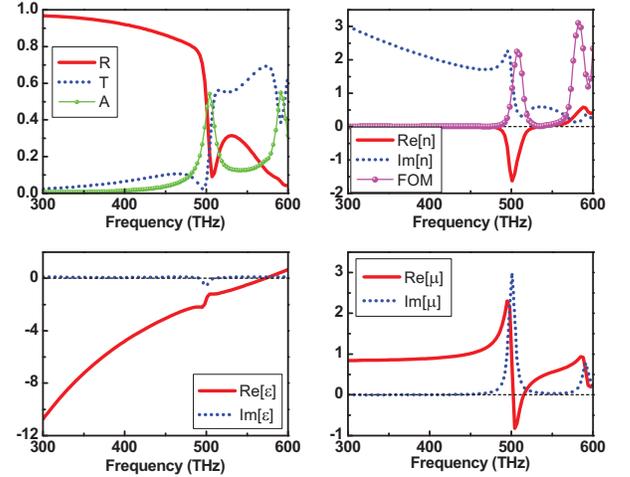}
 \caption{(Color online) Simulated reflection, transmission, absorption, and
retrieved electromagnetic parameters for structure 2, where silver
metal is applied with the corresponding JC data.}
\end{figure}

In conclusion, when MMs go to the optical regime, metallic
inclusions begin to play a crucial role to determine the performance
of MMs. Compared to various Drude models, which may significantly
deviate from the realistic property of the metal for the visible
wavelengths, experimentally-measured JC data will be a better choice
to model metals to provide accurate predictions for MMs' behaviors.
Also, we determined silver, which has a low intrinsic loss, is the
best choice for metallic inclusions of optical MMs for achieving
negative refractive index.

Work at Ames Laboratory was supported by the Department of Energy
(Basic Energy Sciences) under contract No. DE-AC02-07CH11358. This
was partially supported by the European Community Project
NIM\rule[-2pt]{0.2cm}{0.5pt}NIL, Contract No. 228637.

\bibliographystyle{apsrev}

\newpage

\end{document}